\documentclass[12pt]{article}

\begin{document}

\begin{titlepage}

\begin{flushright}
IUHET-493\\
\end{flushright}
\vskip 2.5cm

\begin{center}
{\Large \bf Limits on Lorentz Violation from Synchrotron and Inverse Compton
Sources}
\end{center}

\vspace{1ex}

\begin{center}
{\large B. Altschul\footnote{{\tt baltschu@indiana.edu}}}

\vspace{5mm}
{\sl Department of Physics} \\
{\sl Indiana University} \\
{\sl Bloomington, IN 47405 USA} \\

\end{center}

\vspace{2.5ex}

\medskip

\centerline {\bf Abstract}

\bigskip

We derive new bounds on Lorentz violations in the electron sector
from existing data on high-energy
astrophysical sources. Synchrotron and inverse Compton data give precisely
complementary constraints. The best bound on a specific combination of
electron Lorentz-violating
coefficients is at the $6\times 10^{-20}$ level, and independent bounds are
available for all the Lorentz-violating $c$ coefficients at the
$2\times 10^{-14}$ level or better. This represents an improvement in some bounds
by fourteen orders of magnitude.

\bigskip

\end{titlepage}

\newpage

The possibility that Lorentz invariance may not be exact in nature has been
a subject of great interest since the discovery that Lorentz symmetry could be
broken spontaneously in string theory~\cite{ref-kost18}. Any observed deviations
from Lorentz invariance would be powerful clues about the nature of Planck-scale
physics. There has been a great deal of work probing the physics of Lorentz
violation, both experimental and theoretical.
Sensitive tests of Lorentz symmetry
have included studies of matter-antimatter asymmetries for
trapped charged particles~\cite{ref-bluhm1,ref-bluhm2,ref-gabirelse,
ref-dehmelt1} and bound state systems~\cite{ref-bluhm3,ref-phillips},
determinations of muon properties~\cite{ref-kost8,ref-hughes}, analyses of
the behavior of spin-polarized matter~\cite{ref-kost9,ref-heckel},
frequency standard comparisons~\cite{ref-berglund,ref-kost6,ref-bear,ref-wolf},
Michelson-Morley experiments with cryogenic resonators~\cite{ref-muller1,
ref-stanwix}, Doppler effect measurements~\cite{ref-saathoff,ref-lane1},
measurements of neutral meson
oscillations~\cite{ref-kost10,ref-kost7,ref-hsiung,ref-abe},
polarization measurements on the light from distant galaxies~\cite{ref-carroll1,
ref-carroll2,ref-kost11}, and others.

On the theoretical side, a Lorentz- and CPT-violating effective field theory,
the standard model extension (SME) has been developed in detail~\cite{ref-kost1,
ref-kost2}. Basic issues regarding this theory, including stability and
causality~\cite{ref-kost3} and one-loop renormalizability~\cite{ref-kost4} have
been addressed. The SME contains coefficients that parameterize possible
Lorentz violations. Many of these coefficients
are tightly constrained by the various experiments, but many others are not.

In this paper, we shall provide some further bounds on an important but relatively
poorly constrained sector of the SME. By analyzing data from high-energy
astrophysical sources, we can get strong new constraints, the best of which is at
the $6\times 10^{-20}$ level. We shall use data on both synchrotron
and inverse Compton (IC) emissions; it turns out that these two types of radiation
give complementary bounds.

Synchrotron radiation has previously been used to bound
nonrenormalizable Lorentz-violating parameters~\cite{ref-jacobson1,ref-jacobson2}.
The Crab nebula shows evidence of synchrotron emission from electrons with
Lorentz factors of $\gamma=\left(1-\vec{v}\,^{2}\right)^{-1/2}\sim3\times10^{9}$.
The presence of electrons with velocities this large can be used to constrain
models with deformed dispersion relations. For a nonrenormalizable
Lorentz-violating coefficient
with a particular sign, the data show that the coefficient must be at least
seven orders of magnitude smaller than ${\cal O}(E/M_{P})$ Planck-level
suppression. We shall use a similar technique, but concentrating on the more
important renormalizable Lorentz-violating
operators. We have previously analyzed synchrotron emission in the presence of
these kinds of Lorentz violations and presented bounds based on the data from the
Crab nebula~\cite{ref-altschul5}. There have also been more detailed theoretical
analyses of Lorentz-violating synchrotron processes with nonrenormalizable
operators~\cite{ref-motemayor1} or noncommutative geometry~\cite{ref-castorina}.

We shall consider here a carefully chosen subset of the SME coefficients.
The Lagrange density for our theory is
\begin{equation}
\label{eq-L}
{\cal L}=-\frac{1}{4}F^{\mu\nu}F_{\mu\nu}+
\bar{\psi}[(\gamma^{\mu}+c^{\nu\mu}\gamma_{\nu})
(i\partial_{\mu}-eA_{\mu})-m]\psi,
\end{equation}
where $\psi$ is the electron field.
$c$ contains nine parameters that contribute to Lorentz-violating physics at leading
order. The trace $c^{\mu}\,_{\mu}$ only affects the overall normalization of the
electron
field, and the antisymmetric part of $c$ has no effects at first order, where it
is equivalent to a redefinition of the Dirac matrices. We shall not use a symmetric
$c$, however, but shall instead set the 
$c^{\nu0}$ coefficients to zero using a field redefinition, so that the theory
theory may be straightforwardly quantized and to simply the coupling to $A^{\mu}$.

The SME $c$ coefficients modify the kinetic part of the electron Lagrangian, so they
grow in importance at high energies. Other kinetic modifications
are less important. Most of the operators in question violate
the $SU(2)_{L}$ gauge invariance of the standard model, so they can only arise
as vacuum expectation values of higher dimensional (i.e., nonrenormalizable)
operators; we therefore expect them to be suppressed. There is also a $d$
interaction, which is gauge invariant, similar in form to $c$ but with
an additional $\gamma_{5}$. However, effects from $d$ should average out,
because there should be no net polarization of the electrons in high-energy
sources. Mixing between the standard kinetic term and the $c$ term (which have the
same basic Dirac structure) also causes the
effects of $c$ also grow in importance relative to the effects of the other SME
coefficients at high energies~\cite{ref-kost3}.

The electromagnetic sector in (\ref{eq-L}) is conventional.
Purely electromagnetic violations of Lorentz invariance fall into two
categories. Some terms produce photon birefringence, and all these are very
tightly constrained. The other terms can be absorbed into the $c$ coefficients
through a coordinate redefinition~\cite{ref-kost17}. Therefore, the terms we
consider here are the only ones which should be relevant experimentally.

The $c$ coefficients for nucleons are quite tightly constrained, but those for
electrons are not. However, in high-energy astrophysical processes, electrons
achieve the largest Lorentz factors and can dominate emissions. The processes we
shall consider depend on the ambient electromagnetic field---synchrotron emissions
depending
primarily on the magnetic field strength and IC emissions on the photon
density. As is true conventionally, the electrostatic potential $\Phi=A^{0}$ is
coupled simply to the charge density $e\psi^{\dag}\psi$, and the
vector potential $\vec{A}$ couples to $e\psi^{\dag}\dot{\vec{x}}\psi$.
Radiation from an ultrarelativistic charge is beamed into a narrow pencil of
angles centered around the direction of the charge's velocity. So the radiation
we observe from a distant source comes from electrons that are moving in the
source-to-Earth direction.
The emissions from both processes also show a marked dependence on the relation
that connects energy and velocity, and this is the effect we shall utilize in order
to constrain $c$.

The velocity $\vec{v}$ of an electron is related to its energy $E$ and momentum
$\vec{\pi}$ according to~\cite{ref-altschul4}
\begin{equation}
\label{eq-vfull}
v_{k}=\frac{1}{E+c_{0j}\pi_{j}}\left(\pi_{k}-c_{kj}\pi_{j}-c_{jk}\pi_{j}+c_{jk}c_{jl}
\pi_{l}\right)-c_{0k}.
\end{equation}
Neglecting the terms beyond leading order in $c$ (which should be miniscule),
the maximum velocity in a particular direction described by a unit vector
$\hat{e}$ is $1-c_{jk}\hat{e}_{j}\hat{e}_{k}-c_{0j}\hat{e}_{j}$. With Lorentz
violation, this can be
less than one. If a body's synchrotron
spectrum indicates the presence of electrons with Lorentz factors up
to at least some value $\gamma_{\max}$, then the Lorentz-violating coefficients for
the electron must be bounded according to
\begin{equation}
\label{eq-synclimit}
c_{jk}\hat{e}_{j}\hat{e}_{k}+c_{0j}\hat{e}_{j}<\frac{1}{2\gamma_{\max}^{2}},
\end{equation}
where $\hat{e}$ is in the source-to-Earth
direction just mentioned. [By the Lorentz
factor $\gamma$, we mean precisely the quantity $\gamma=(1-\vec{v}\,^{2})^{-1/2}$.]
This is a one-sided limit, so only coefficients with particular signs are
restricted. In particular, we can only find upper limits on the diagonal components
of $c$ this way; no negative values of $c_{jk}$ for $j=k$ can be excluded.
So a complementary measurement is needed if the
Lorentz-violating coefficients are to be restricted to a bounded region of
the parameter space.

Fortunately, complementary bounds are possible. Just as a positive
$c_{jk}\hat{e}_{j}\hat{e}_{k}+c_{0j}\hat{e}_{j}$ leads to a maximum electron speed
that is less
than one, a negative value for $c_{jk}\hat{e}_{j}\hat{e}_{k}+c_{0j}\hat{e}_{j}$ can
indicate
a maximum electron speed that is greater than one. However, for electrons that
are coupled conventionally to the electromagnetic field, new effects would come
into play if the speeds actually became superluminal. The validity of the model
is questionable under these circumstances, so we cannot make use of the maximum
velocity directly. On the other hand, there is instead a maximal energy attainable
by subluminal electrons, and we can use this to set further bounds on $c$.

In IC scattering, an ultrarelativistic electron scatters off a
comparatively low-energy photon, transferring a substantial fraction of its energy
to the photon. For kinematical reasons, the emitted radiation is (as with synchrotron
radiation) beamed into a narrow pencil of angles centered around the direction of
the electron's motion. The resulting radiation can be observed and used to bound
the Lorentz violation. The key is that the initial electron must have a greater
energy than the observed IC $\gamma$-ray. If the emission from a
source is well modeled by known processes, we may infer that superluminal electrons
(which would produce vacuum Cerenkov radiation and emit synchrotron radiation
at a rate which diverges in the approximation of no
back reaction) are not the source of the emission. Then the
energies of the most energetic IC photons represent lower bounds on the energies
of subluminal electrons; and if there are such electrons up to some energy
$E_{\max}$, then the $c$ coefficients must satisfy
\begin{equation}
\label{eq-IClimit}
-c_{jk}\hat{e}_{j}\hat{e}_{k}-c_{0j}\hat{e}_{j}<\frac{1}{2(E_{\max}/m)^{2}},
\end{equation}
since otherwise, an electron with energy $E_{\max}$ moving in the Earthward
direction would be superluminal.

The similarity between equations (\ref{eq-synclimit}) and (\ref{eq-IClimit}) is
not coincidental. In each case, the effect we want to measure is a product of how
the Lorentz violation affects the relationship between velocity and energy.
In the Lorentz-invariant theory, $\gamma=E/m$,
and by measuring $E$ and $\gamma$
separately, we can get bounds on any deviations from Lorentz invariance.

There are nine components of $c$ that can be bounded in this way---the three
$c_{0j}$ and the six-component symmetric part of $c_{jk}$. If each of these
coefficients is to be bounded on both sides, we must obtain at least
ten inequalities of the forms (\ref{eq-synclimit}) or (\ref{eq-IClimit}),
corresponding to emissions from at least nine separate sources.
Each of these inequalities generally couples all nine of the coefficients in a
nontrivial way, but they may be translated into separate
limits on individual coefficients by means of linear programming.

\begin{table}
\begin{center}
\begin{tabular}{|l|c|c|c|c|c|}
\hline
Emission source & $\hat{e}_{X}$ & $\hat{e}_{Y}$ & $\hat{e}_{Z}$ &
$\gamma_{\max}$ & $E_{\max}/m$ \\
\hline
3C 273 & 0.99 & 0.13 & $-0.04$ &
$3\times10^{7}$\cite{ref-roser} & $2\times10^{5}$\cite{ref-roser} \\
Centaurus A & 0.68 & 0.27 & 0.68 & $2\times 10^{8}$\cite{ref-kataoka} & - \\
Crab nebula & $-0.10$ & $-0.92$ & $-0.37$ & $3\times
10^{9}$\cite{ref-aharonian1} & $2\times 10^{8}$\cite{ref-aharonian1} \\
G 12.82-0.02 & $-0.06$ & 0.95 & 0.29 &
- & $5\times 10^{7}$\cite{ref-aharonian6} \\
G 347.3-0.5 & 0.16 & 0.75 & 0.64 &
$3\times 10^{7}$\cite{ref-ellison} & $2\times 10^{7}$\cite{ref-aharonian3} \\
MSH 15-52 & 0.34 & 0.38 & 0.86 &
- & $8\times 10^{7}$\cite{ref-aharonian4} \\
PSR B1259-63 & 0.42 & 0.12 & 0.90 &
- & $6\times 10^{6}$\cite{ref-aharonian5} \\
RCW 86 & 0.35 & 0.30 & 0.89 &
$10^{8}$\cite{ref-rho} & - \\
SNR 1006 AD & 0.52  & 0.53 & 0.67 &
$2\times 10^7$\cite{ref-allen} & $7\times 10^{6}$\cite{ref-allen} \\
Vela SNR & 0.44 & $-0.55$ & 0.71 & $3\times 10^{8}$\cite{ref-mangano} &
$1.3\times 10^{8}$\cite{ref-aharonian2} \\
\hline
\end{tabular}
\caption{
\label{table-combined}
Parameters for the astrophysical sources that we shall use to constrain $c$.
The coordinates $X$, $Y$, and $Z$ are in sun-centered celestial equatorial
coordinates~\cite{ref-bluhm4}.
References are given for each value of
$\gamma_{\max}$ or $E_{\max}$.}
\end{center}
\end{table}

Table~\ref{table-combined} lists the parameters for ten astrophysical sources for
which useful bounds are available. Most of the sources are supernova remnants, but
a few extragalactic sources are included. The radio galaxy Centaurus A and the
quasar 3C 273 are among the brightest and best understood objects of their
kinds. All the sources can be modeled relatively cleanly, giving fairly secure
values of $\gamma_{\max}$ and/or $E_{\max}$. The $\gamma_{\max}$ values must be
extracted directly from the models, and so errors in the modeling could affect the
validity of the limits; this makes it important that only well-understood sources be
used. The $E_{\max}$ limits are less model dependent. IC models of $\gamma$-ray
sources typically require maximum electron energies that are several times larger
than the highest observed photon energies (because an IC scattering event does not
transfer all of the electron's energy to the photon). However, this model-derived
$E_{\max}$ is not the value we have chosen to use in table~\ref{table-combined}.
Instead, we have conservatively identified the highest actually observed photon
energy as $E_{\max}$; the only input from a model is that the source's
$\gamma$-ray emission is well described by the IC process. This choice of $E_{\max}$
ensures that Lorentz-violating distortions of the energy-momentum relation at
higher than observed energies (which could invalidate the model results) are not a
problem.

The bounds derived from the parameters in table~\ref{table-combined} (the best of
which is at the $6\times10^{-20}$ level) are slightly awkward. It is difficult
with nine coupled parameters to determine exactly what regions of the parameter
space are being excluded. We have therefore extracted, via linear programming,
independent limits on each of the nine components of $c$. These bounds represent the
absolute maximum and minimum values that are possible for each coefficient. They
are therefore not as tight numerically as the raw bounds.

In addition to the astrophysical bounds derived here, we have included some other,
roughly comparable bounds in the linear program. Optical resonator tests are usually
used to place bounds on the parameters of the SME photon sector. However, these
same experiments may be used to place bounds on the electron $c$
coefficients~\cite{ref-muller2}. The key realization is that electronic Lorentz
violations will modify the structure of a crystalline resonator, and this effect
can be worked out systematically, provided Lorentz violations for nucleons can be
safely neglected.
The result is further bounds on $|c_{(XY)}|$, 
$|c_{(XZ)}|$, $|c_{(YZ)}|$, and $|c_{XX}-c_{YY}|$, which we shall all conservatively
take to be at the $3\times 10^{-15}$ level.
[Here, $c_{(jk)}$ means the symmetric sum $c_{jk}+c_{kj}$.]

\begin{table}
\begin{center}
\begin{tabular}{|c|c|c|}
\hline
$c_{\mu\nu}$ & Maximum & Minimum \\
\hline
$c_{XX}$ & $5\times 10^{-15}$ & $-6\times 10^{-15}$ \\
$c_{YY}$ & $3\times 10^{-15}$ & $-3\times 10^{-15}$ \\
$c_{ZZ}$ & $5\times 10^{-15}$ & $-3\times 10^{-15}$ \\
$c_{(XY)}$ & $3\times 10^{-15}$ & $-3\times 10^{-15}$ \\
$c_{(YZ)}$ & $3\times 10^{-15}$ & $-3\times 10^{-15}$ \\
$c_{(YZ)}$ & $3\times 10^{-15}$ & $-3\times 10^{-15}$ \\
$c_{0X}$ & $5\times 10^{-15}$ & $-2\times 10^{-14}$ \\
$c_{0Y}$ & $2\times 10^{-15}$ & $-5\times 10^{-16}$ \\
$c_{0Z}$ & $10^{-16}$ & $-5\times 10^{-17}$ \\
\hline
\end{tabular}
\caption{
\label{table-separate}
Independent bounds on the components of $c$.}
\end{center}
\end{table}

The results are given in table~\ref{table-separate}. The cryogenic resonator bounds
are still the best for $|c_{(XY)}|$,  $|c_{(XZ)}|$, and $|c_{(YZ)}|$, but bounds
on almost all the other coefficients are now comparable or better. The bounds on
$c_{0Z}$ are especially strong, while in laboratory experiments,
boost invariance violation coefficients such as $c_{0j}$ are typically harder to
constrain. This shows the advantage of deriving bounds from emissions by
relativistic sources. The bounds on the separate coefficients are
completely independent; the maximum and minimum values presented are the largest and
smallest that a given coefficient can take under any circumstances. Moreover, there
are also additional correlations; it is not generally possible for several of the
coefficients to take on their extremum values simultaneously.

This method for placing bounds on Lorentz violation could not be used directly to
identify actual Lorentz violation. The signature of Lorentz violation in
high-energy astrophysical sources would be emission spectra that cannot be modeled
by conventional radiation mechanisms. If the calculated bounds on certain
Lorentz-violating coefficients were significantly weaker than others, that would be
an indication that those coefficients might actually be nonzero. However, since most
of the bounds in table~\ref{table-separate} are comparable in magnitude, we see no
indications that any particular components of $c_{\mu\nu}$ are more likely to be
nonzero than others.

Several of the bounds we have derived are much better than
previous constraints. The previous bounds on the electronic $c_{0j}$
coefficients were at the $10^{-2}$ level~\cite{ref-lane1}, so the improvements here
are by more than
eleven orders of magnitude (fourteen orders for $c_{0Z}$). Moreover, all nine of
of the coefficients that contribute at leading order can be comparably bounded by
this method, and as better measurements from a wider variety of astrophysical
sources become available, the limits on $c$ will improve as well.

%
%
%
%
%
%
%
%
%
%
%
%

\section*{Acknowledgments}
The author is grateful to V. A. Kosteleck\'{y} and Q. G. Bailey for helpful
discussions.
This work is supported in part by funds provided by the U. S.
Department of Energy (D.O.E.) under cooperative research agreement
DE-FG02-91ER40661.

\end{document}